# Spintronic Switches for Ultra Low Energy On-Chip and Inter-Chip Current-Mode Interconnects

Mrigank Sharad and Kaushik Roy, Fellow, IEEE
Department of Electrical and Computer Engineering, Purdue University, West Lafayette, IN, USA
{msharad, kaushik}@.purdue.edu

*Abstract*— Energy-efficiency and design-complexity of high-speed on-chip and inter-chip data-interconnects has emerged as the major bottleneck for high-performance computing-systems. As a solution, we propose an ultra-low energy interconnect design-scheme using nano-scale spin-torque switches. In the proposed method, data is transmitted in the form of current-pulses, with amplitude of the order of few micro-amperes that flows across a small terminal-voltage of less than 50mV. Sub-nanosecond spin-torque switching of scaled nano-magnets can be used to receive and convert such high-speed current-mode signal into binary voltage-levels using magnetic-tunnel-junction (MTJ), with the help of simple CMOS inverter. As a result of low-voltage, low-current signaling and minimal signal-conversion overhead, the proposed technique can facilitate highly compact and simplified designs for multi-gigahertz inter-chip and on-chip data-communication links. Such links can achieve more than ~100x higher energy-efficiency, as compared to state of the art CMOS interconnects.

*Index Terms*—Integrated circuit interconnections, spin valves, magnets.

## I. INTRODUCTION

The ever increasing demand for higher computing-capabilities has necessitated the integration of multiple processing cores and larger memory-blocks, resulting in increasingly busy inter-chip links and complex, power-hungry input/output (I/O) interfaces for microprocessors [1]. The same is true with respect to on-chip global interconnects like, multi-byte buses for on-chip memory-read and long-distance inter-block links [2]. Moreover, with the scaling of the CMOS technology, energy efficiency and performance of the on-chip global-interconnect degrades due to increase in per-unit length resistance of long metal-lines [2]. As a result, the design of inter-chip and on-chip global interconnects has emerged as a major challenge for large-scale parallel computing systems.

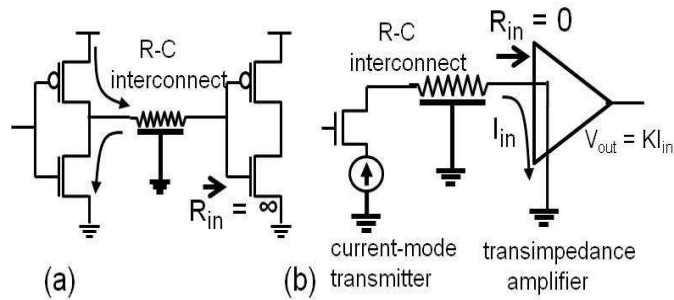

Fig. 1(a) Voltage-mode interconnects that involves capacitive switching and offers high input impedance to the link (b) current-mode interconnect with a low input-impedance receiver.

Solutions at technology, circuit, and system level have been explored to address the aforementioned design challenges pertaining to interconnects [1-8]. For instance, the use of current-mode signaling for long distance links has been shown to offer reduced power consumption and enhanced bandwidth [4, 21] (fig. 1). This is because current-mode transmission reduces the voltage swing on the metal-lines, thereby reducing the capacitive switching power. Also, the receiver for current-mode links are designed to provide minimal input impedance to the transmission line (as opposed to voltage-mode links, which provide high impedance capacitive-load). This results in higher bandwidth, as compared to voltage-mode signaling. Increased bandwidth alleviates the need of equalization at the receiver end to a significant extent. However, analog-based current-mode transceivers are more complex than simple inverters, used for voltage-mode links, and add significantly to static-power consumption as well as area complexity, at the I/O interfaces [4, 5]. As a technology solution, use of optical interconnects for inter-chip [6, 7] as well as on-chip data links [8], has been proposed. But again, optical modulators (at the transmitting side) and receivers consume large amount of power and area that can eschew their overall benefits [8].

In this work we propose an alternate technology solution that can potentially lead to ultra-low energy, high-speed data links with highly simplified I/O interfaces. Recent experiments have shown that spin-polarity of nano-scale magnets can be flipped at sub-nanosecond speed using small charge currents [9-13]. Application of current-induced spin-torque switching of nano-magnets for memory and logic-design has been proposed in literature [12, 17]. In this work we explore the possibility of applying such nano-scale spin-torque switches to the design of ultra low-voltage, current-mode on-chip and inter-chip transmission links. Magneto-metallic spin-torque devices, like domain-wall magnet [9], and spin-valves [14], can act as ideal receivers for current-mode signals, owing to their small resistance and the possibility of low-current high-speed switching [14, 15]. Such low resistance receiver ports can allow ultra-low voltage biasing of the entire communication link, thereby reducing the static power consumption due to current-mode signaling. Moreover such devices can facilitate easy conversion of current-mode signal into full-swing on-chip voltage levels, through the use of magnetic tunnel junctions (MTJ) [17]. As a result, nano-scale spin torque devices can lead to very compact and highly energy-efficient on-chip and inter-chip interconnects for large-scale parallel-computing systems.

Rest of the paper is organized as follows. Section 2 presents the basic spin-torque device structure that can be used for high-speed interconnect design. Section 3 describes the circuit design for on-chip and inter-chip links using the spin-torque device. Section 4 briefly discusses the prospects of the proposed design. An alternate spin-torque based device option for high-speed current-mode interconnect is discussed in section. 5. Simulation-framework is briefly described given in section. 6. Conclusions are given in section 7.

## II. SPIN-TORQUE SWITCH FOR CURRENT MODE INTERCONNECTS

In this section we describe spin-torque device structures suitable for the design of high speed current-mode links. We apply domain-wall shift based switch in this work, however, a brief discussion on the applicability of other spin-torque devices is given towards the end.

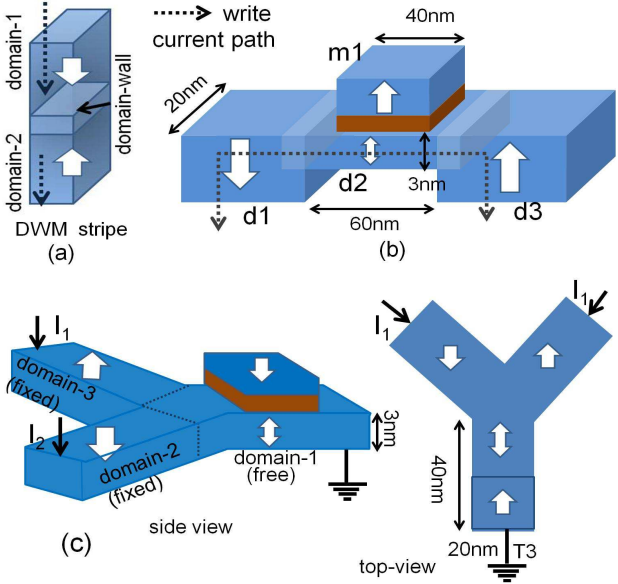

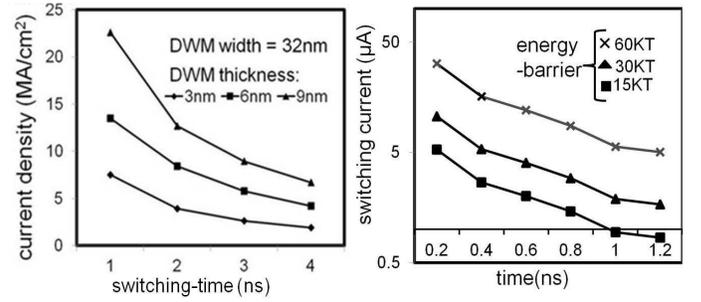

Fig. 3. (a) Plots for switching current density vs. switching time for different thickness of $d_2$, showing that device-scaling can reduce the critical current-density for a given switching time, higher current can be used for faster switching, (b) Lower anisotropy energy barrier can be exploited for high-speed low-current switching.

Fig. 2 (a) A domain-wall-magnet (NiFe) with three magnetic domains, (b) Unipolar Domain Wall Switch (UDWS)[12]. , (c) Bipolar Domain Wall Switch (BDWS) [17].

A domain-wall-magnet (DWM) constitutes of multiple *nano-magnet* domains separated by non-magnetic regions called domain-wall (DW) as shown in fig. 2a. DW can be moved along a magnetic nano-strip using current-injection. Hence, the spin-polarity of the DWM strip at a given location can be switched, depending upon the polarity of its adjacent domains and the direction of current-flow. Experiments have achieved switching current density of the order of $10^6 A/cm^2$ for nano-scale DWM strips, and, a switching time of less than 500ps [9]. Thus, the polarity of a scaled *nano-magnet* strip of dimension $3x20x60 nm^3$ can be switched using a small current of ~1μA.

The device structure for a unipolar domain-wall-switch (DWS), is shown in fig. 2b. It constitutes of a thin and short ($3x20x60 nm^3$) *nano-magnet* domain, $d2$ (domain-2, the 'free-domain') connecting two anti-parallel *nano-magnet* domains of fixed polarity, $d1$ (domain-1) and $d3$ (domain-3) (fixed through exchange coupling to larger magnets [12]). Domain-1 forms the input port, whereas, domain-3 is grounded. Spin-polarity of the free-domain ($d2$) can be written parallel to $d1$ or $d3$ by injecting a small current along it from $d1$ to $d3$ and vice-versa. Thus, the DWN can detect the polarity of the current flow at its input node. Hence, it acts as an ultra-low-voltage and compact current comparator [17], that can be employed for recovering data from a bipolar, current-mode signal received at its input.

To achieve maximum possible energy efficiency, it is desirable to minimize the switching current threshold for the DWS. Notably, scaling the device-dimensions reduces the critical switching-current, as demonstrated in experiments [12]. Similar trends are depicted in the micro-magnetic simulation plots shown in fig. 3a. Apart from device-scaling, the use lower anisotropy barrier for the magnetic material can be effective in lowering the switching threshold for high speed computing applications [19] (fig. 3b). A DWS with an anisotropy barrier of 15KT and dimensions given in fig. 1b, can possibly achieve a switching speed of ~100ps with an input current magnitude ~30μA (fig.3b).

Fig. 4b shows an alternate device structure based on domain wall magnet, bipolar domain wall switch [17]. It constitutes of two fixed-domains of opposite polarity (domain-2 and domain-3) that act as input ports and polarize the input currents. The third domain (domain-3) is the switchable free-domain. The spin polarity of the current injected into the free domain is effectively the difference between the current inputs $I_1$ and $I_2$ entering through the two free domains. Hence depending upon which of the two input is larger, the free domain can switch parallel to either of the two fixed input domains. Hence, similar to unipolar-DWS discussed above, previous section, this device acts as a current-comparator. The minimum difference between the two input currents the bipolar-DWS (BDWS) can detect would depend upon the critical current density for domain-wall shift in the free domain. For a $20x2$ $nm^2$ cross-section area of domains, and a critical current density of the order of $10^6$ $A/cm^2$, a device with relaxed energy barrier (10KT), can possibly detect a difference of ~1μA. The state of free domain (domain-3) can be read using the MTJ formed on its top.

Two complementary input ports for the BDWS can facilitate simplified circuit designs as will shown in the following sections. However, the magnitude of input currents that this device can handle might be smaller. This is because, for the BDWS, the sum of the two charge current inputs (I1 and I2) enter into the device and the subtraction in achieved through spin-polarization. However, for the UDWS, the difference between two currents to be compared, can be injected into the device, thereby reducing the charge current magnitude flowing through the free domain [17].

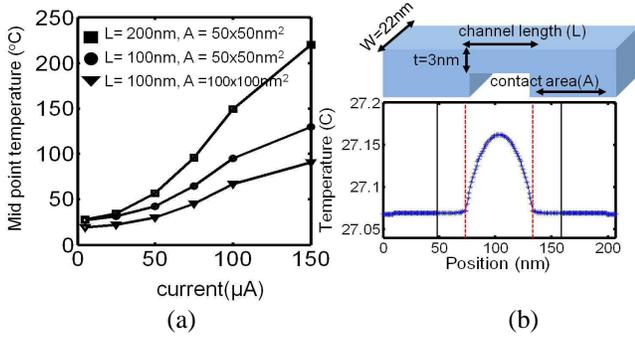

Fig. 4 (a) COMSOL simulation for temperature rise in the DWN device for different device dimensions, (b) plot showing temperature profile along the device for a small input current of ~1μA.

The upper limit upon the permissible current density and hence upon the switching speed may be determined by the Joule-heating effect in the DWS. The effect of Joule heating in the device was simulated using finite-element simulation through COMSOL [20]. The thin and short central free-domain of the device is the most critical portion with respect to current driven heating (fig.4a). Plot in fig. 4 shows that the heating in the device can be reduced by choosing larger contact area of the two fixed domains. Also, shorter free domain results in smaller heating. Thus, the current handling capacity of the device can be increased by appropriate structural optimization.

In order to read the state of the free domain *d2* of the DWS, a magnetic tunnel junction (MTJ) [17], formed between a fixed polarity magnet *m1* and *d2* is employed. The effective resistance of the MTJ is smaller when *m1* and *d2* have the same spin-polarity and vice-versa. A large ratio between these two resistance state, defined in terms of tunnel-magnetoresistance-ratio (TMR) [17], can facilitate simplified read operation. A compact and low power CMOS circuit like a differential MTJ latch [16, 17] or even a simple CMOS inverter can be employed to convert the spin-mode information received by the DWS into binary voltage levels. In the following section we describe a possible circuit realization for high-speed I/O links using the DWS device.

### III. INTERCONNECT DESIGN USING DWS

Fig. 5 depicts the circuit for a generic high-speed data interconnect employing a DWS-based receiver. At the transmitter end, the PMOS transistors M1 and M2 are driven by voltage-mode data-signal. Their source terminals are biased at two different DC-voltages V+ΔV and V-ΔV, where V is 0.7V and ΔV can be ~25mV. On the receiver side, the DWS is biased at a voltage V, as shown in the figure. This effectively biases the transistors M1 and M2 across a small drain to source-voltage ΔV-δV, where δV is the voltage drop across the transmission line. Thus, M1 and M2 operate in linear region and act as deep-triode current-sources (DTCS). Depending upon the value of the voltage-mode data signal at the transmitter, either M1 or M2 are turned ON, resulting in either positive or negative current flow across the DWS at the receiver. This results in data-dependent flipping of the DWS free-layer, which can be detected using a resistive voltage divider formed with a reference MTJ as show in fig. 5a. A high TMR for the MTJ can provide higher voltage swing close to V/2, where V is the inverter supply voltage. The output of the voltage divider can be sensed by a simple CMOS inverter.

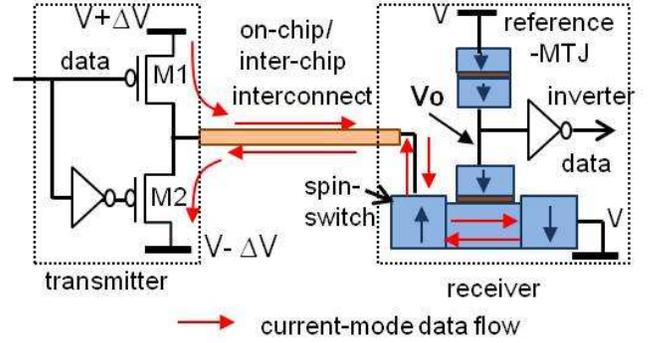

Fig. 5a Circuit for on-chip and inter-chip interconnect using UDWS

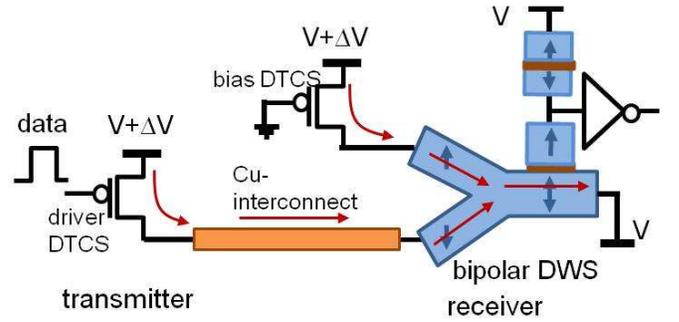

Fig. 5b Circuit for on-chip and inter-chip interconnect using BDWS

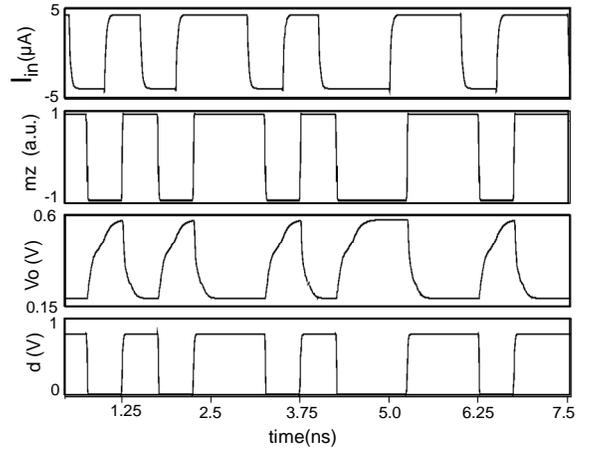

Fig. 6 Simulation plots for 2Gbps signaling over 2mm long on-chip interconnect.

Tunneling current in the voltage divider needs to be minimized in order to reduce the static power consumption and to avoid disturbance of the free-layer state. This requirement is easily met by employing higher oxide-thickness, $t_{ox}$, for the MTJ. However, higher $t_{ox}$ value leads to higher resistance and hence slower signal transition at the output-node, due to R-C delay. Hence, an optimal value of $t_{ox}$ can be chosen, depending upon the data transmission frequency. Differential CMOS latch can also be employed to avoid static current-path [17]. Fig. 6 shows the plot for 2Gbps signaling over a 2mm long on-chip interconnect using the proposed scheme.

The transceiver circuit realization using bipolar DWS (BDWS) is also shown in fig. 5b. In this case, a constant bias current is injected into one of the inputs of the BDWS on the receiver side. The DTCS at the transmitter injects current on the transmission line only when data value is '1'. This current is larger (by a factor of ~2X) than that provided by the bias DTCS at the receiver, and is received by the second input of the BDWS. Thus, the DWS free layer at the receiver is switched depending upon the data value transmitted. The BDWS facilitates relatively simpler circuit realization as it requires only one extra voltage level ($V+\Delta V$) apart from a main supply of value $V$.

The interconnect schemes depicted in fig. 5 can also extended to the design of on-chip clock-synchronized global interconnects, as shown in fig. 7 [2]. It constitutes of compact spin-CMOS hybrid repeaters that can propagate low-voltage, current-mode signal across multi-bit buses (fig. 8).

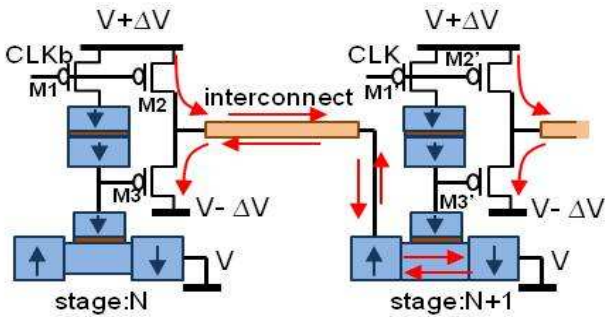

Fig. 7 Clock-synchronized on-chip interconnect with DWS-based repeaters.

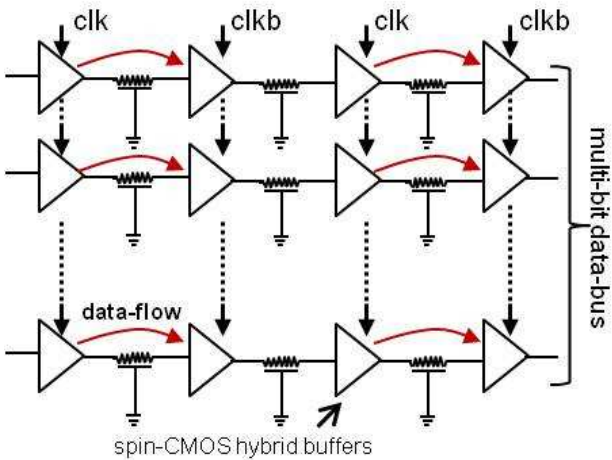

Fig. 8 clock-synchronized multi-bit data bus using spin-CMOS hybrid current-mode interconnects.

At the transmitter side, the transistor M3 is conditionally put *ON* depending upon the spin-mode data stored in the corresponding DWS. The transistor M2 on the other hand is clock-driven and effectively subtracts a net current from that supplied by M3. Thus, when the clock for the $N_{th}$ stage turns low, its data is transferred to the DWS of the $N+1_{th}$ stage through current-mode signal. Such a scheme can be employed to design compact and energy-efficient current-mode repeaters for long-distance global interconnects for high performance microprocessors [2].

## IV. PERFORMANCE AND PROSPECTS

In the proposed interconnect-design scheme, the energy consumption per-bit transmitted can be evaluated as the sum of static power dissipation across the transmission line $E_{int}$, the power consumption in the conversion circuit $E_{conv}$, at the receiver. The dynamic switching power for the small-size digital driver at the transmitter can be negligibly small as compared to the aforementioned components. Similarly, the switched-capacitance power consumption on the link is relatively small because of current-mode signaling and very small voltage fluctuation on the metal lines (close to ~10mv).

The DWS facilitates ultra-low voltage biasing of the entire transmission link, such that the static current flows across a small terminal voltage of $2\Delta V$. For a 10mm long on-chip interconnect (parameters given in [4]) would offer a resistance of ~500Ω and an effective capacitance of ~2.5pF. Transmission of data at 2Gbps (data-period $T_d$ =0.5 ns) speed over such a link may require a current-amplitude ($I_d$) of ~10μA in order to be able to switch the DWS. This current magnitude can be supplied by minimum size transistors M1 and M2 (with effective resistance of ~1kΩ in 45nm CMOS technology) with a $\Delta V$ of ~25mV. The component $E_{int}$ can be therefore calculated as $E_{int} = 2\Delta V \times I_d \times T_d$, which evaluates to ~0.25fJ. The power consumption in the detection unit can be minimized with the optimal choice of $t_{ox}$ as discussed earlier. A TMR of 400% was used for the MTJs. For 2Gbps operation, the power consumption in the optimized detection circuit was found to be ~0.8μW. This translates to a value of ~0.4fJ for $E_{conv}$. Thus, the overall energy dissipated per-bit can be ~0.65fJ which is around two order of magnitude less than that reported in a recent mixed-signal CMOS implementation. Moreover, it is evident that the proposed spin-CMOS hybrid interconnect can be very compact and area-efficient as compared to conventional mixed signal CMOS current-mode I/O interfaces. Thus the spin-torque based I/O interfaces can emerge as a very attractive solution to the design challenges associated with on-chip and inter-chip interconnects.

## V. OTHER DEVICE OPTIONS

Other spintronic switches can also be employed in the proposed scheme. For instance, low-current high-speed switching of nano-magnets in lateral spin valves may be facilitated by current-mode Bennett-clocking (Note that current-mode Bennett-clocking has been theoretically proposed in [14], but is yet to be demonstrated experimentally) proposed for All Spin Logic (ASL) devices [14-16, 22]. Such a device is shown in fig. 9.

It constitutes of an output-magnet m1 connected to a fixed input magnet m2 through a copper channel. Another fixed magnet m3, with a spin-polarity orthogonal to that of m1, is used as an assist for switching m1. In the beginning of the data cycle, a relatively large current $I_h$ (~100μA), is injected through m3 into m1, which can reduce the required current threshold for m1 by a large factor (~10). As a result, a small input current $I_{in}$ received through m2 can flip m1 back to one of its stable state (in-plane or out of plane, depending upon direction of current-flow). The use of spin-torque as an assist can be termed as current-mode Bennett clocking [14, 15]. The state spin-state of m1 can be detected with the help of an MTJ formed with a fixed magnet m4, which covers a part of the top area of m1. For small dimensions of m1 (as depicted in the transient plot in fig. 9), a small current of ~10μA received trough a current-mode transmission channel can suffice for inducing high speed switching (~100ps) with the help of current-mode spin-torque assist. This reduces voltage-drop and power consumption in the transmission line, at a small energy-cost of $I_h$ (which is injected through m3, using a small voltage).

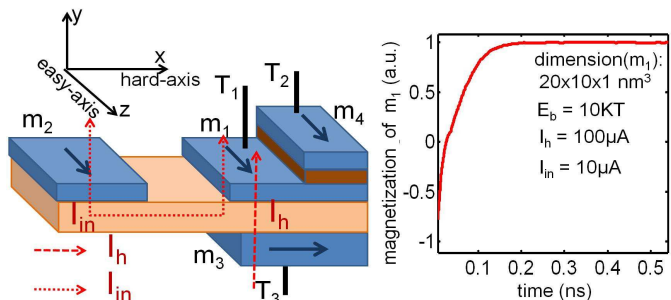

Fig. 9 ASL switch with current-mode Bennett-clocking [14, 15], for high-speed, energy efficient current-mode interconnects

## VI. SIMULATION FRAMEWORK

We employed device-circuit co-simulation framework to simulate the proposed interconnect scheme. The DWS was modeled using physics based micro-magnetic model for domain-wall-magnet in [18] that has been calibrated with experimental data on current-induced domain wall motion. The MTJ characteristic we modeled using self-consistent simulation of Non-equilibrium Greens' Function (NEGF) and Landau-Lifshitz-Gilbert equations (LLG). The two models were solved self-consistently in order to ensure negligible impact of read current through the MTJ's upon the DWS switching dynamics. Behavioral model of the DWS-device derived from the physics-based equations were employed in HSPICE simulation for evaluating the circuit level performance.

## VII. CONCLUSION

We proposed a novel technology solution for on-chip and inter-chip interconnect design using spin-torque switches. Magneto-metallic spin-torque switches act as ideal, low-impedance current-mode receivers, allow ultra low-voltage biasing of the I/O interconnect and facilitate easy conversion from spin to charge using an MTJ interface. As a result, the proposed technique for high-speed current-mode interconnect design can be highly compact and more than two orders of magnitude energy-efficient and, as compared to state of the art technology solutions for on-chip (global) and inter-chip data links. The proposed technique can provide an attractive technology solution to the inter-connect bottleneck faced by large-scale parallel computing systems.